\documentclass[aps,twocolumn,nofootinbib]{revtex4}
\usepackage{amsmath, amsthm, amsfonts, amssymb}
\usepackage{srcltx, dsfont}
\usepackage{graphicx}

\begin{document}

\title{Vacuum Brans-Dicke theory in the Jordan and Einstein frames: can they be distinguished by lensing?}
\author{R.N. Izmailov}
\email{izmailov.ramil@gmail.com}
\affiliation{Zel'dovich International Center for Astrophysics, Bashkir State Pedagogical University, 3A, October Revolution Street, Ufa 450008, RB, Russia}
\author{R.Kh. Karimov}
\email{karimov\_ramis\_92@mail.ru}
\affiliation{Zel'dovich International Center for Astrophysics, Bashkir State Pedagogical University, 3A, October Revolution Street, Ufa 450008, RB, Russia}
\author{A.A. Potapov}
\email{a.a.potapov@strbsu.ru}
\affiliation{Department of Physics \& Astronomy, Bashkir State University, 49, Lenin street, Sterlitamak 453103, RB, Russia}
\author{K.K. Nandi}
\email{kamalnandi1952@yahoo.co.in}
\affiliation{Zel'dovich International Center for Astrophysics, Bashkir State Pedagogical University, 3A, October Revolution Street, Ufa 450008, RB, Russia}
\affiliation{High Energy Cosmic Ray Research Center, University of North Bengal, Darjeeling 734 013, WB, India}

\begin{abstract}
Vacuum Brans-Dicke theory can be self-consistently described in two frames, the Jordan frame (JF) and the conformally rescaled Einstein frame (EF), the transformations providing an easy passage from one frame to the other at the level of actions and solutions. Despite this, the conformal frames are inequivalent describing different geometries. It is shown that the predictions of the weak field lensing (WFL) observables in the EF are different from those recently obtained in the JF for the vacuum Brans-Dicke class 1 solution. The value of the Brans-Dicke coupling parameter $\omega$ from the Cassini spacecraft experiment reveals the degree of accuracy needed to experimentally distinguish the WFL measurements including the total magnification factor in the two frames.
\end{abstract}

\maketitle

\section{Introduction}
Brans-Dicke (BD) theory [1-3], which describes gravitation through a
spacetime metric ($g_{\mu \nu }$), a massless scalar field ($\varphi $) and
a constant coupling parameter $\omega $, is a Machian competitor of
Einstein's general relativity (GR). The theory continues to receive
widespread attention since it is consistent with the solar and cosmological
observations [4-12]. The vacuum BD theory that we shall consider here
remarkably arises also in the low energy limit of heterotic string theory in
4-dimensions that predicts a model independent value $\omega =-1$ [13,14].
The 1-parameter group of conformal symmetry of vacuum BD action discovered
by Faraoni [7] can shift this value to any desired higher value of $\omega $
required by solar system observations.

A generic aspect of any scalar-tensor theory is that two frames are
available for its description. One frame is the so-called Jordan frame (JF)
in which the BD field equations were originally written and the scalar field
$\varphi $ plays the role of a spin-$0$ component of gravity introduced to
accommodate Mach's principle. The other is the conformally rescaled Einstein
frame (EF) in which the transformed scalar field $\phi $ plays the role of
material source stresses in the GR field equations. The curvature and
geodesic properties are different in the two frames. Because of this, there
have been a long standing debate as to which of the two frames, JF and EF,
should be considered physical.\footnote{%
A physical frame is the one in which test particles have constant masses and
move along the geodesics in that frame. Vacuum BD theory in JF ($T_{\mu \nu
}^{\text{(matter)}}=0$) avoids anomalous coupling to matter in the EF. Then
the EF solution is just the scalar field solution of GR corresponding to the
vacuum solution in JF [29]. In this sense, both the frames are physical but
observable predictions can still differ, which is the object of this paper.}

Depending on the attitude to this question, physicists are divided roughly
into six groups. Some authors: (1) neglect the issue, (2) think that the two
frames are physically equivalent, (3) consider them physically
non-equivalent but do not provide supporting arguments, (4) regard only JF
as physical but EF can be used for mathematical convenience, (5) regard only
EF as physical, (6) belong to two or more of the above categories! For a
detailed account, see the excellent review [9]. There are also arguments
claiming that all physical\textit{\ }observables are conformal frame
invariants [15]. Some works in cosmology do show that it is indeed the case
[16,17]. In this situation, we think that it is imperative to go beyond
theoretical arguments favoring one position or the other but analyze
tangible experimental predictions in the two frames. There exist very few
works in this direction. One study involves the response of a gravitational
wave detector to scalar waves, but the distinctive feature of a longitudinal
mode, to be present in EF and absent in JF, seems unlikely to be observed in
the near future [18].

In this paper, we consider the more pragmatic case of weak field lensing
(WFL) in the EF and compare them with those in the vacuum BD theory in the
JF recently obtained by Gao et al [19]. Even though the GR theory in the EF
has nothing to do with the Machian coupling parameter $\omega $, we show
that the EF WFL observables can still be expressed\textit{\ }as functions of
$\omega $ making them amenable to the desired comparison. We shall then
translate the value of $\omega $ obtained from the Cassini spacecraft
experiment [20] into an accuracy level needed to distinguish the two frames.

In Sec.2, we state the solutions in JF and EF with the relation connecting
the respective parameters. In Sec.3, we calculate the WFL observables for
SgrA* and evaluate in Sec.4 the accuracies needed to observationally
distinguish the two frames. Sec.5 discusses a useful WFL observable, the
total magnification. Sec.6 concludes the paper. We take $G=1$, $c=1$ unless
specifically restored.

\section{JF and EF}
\bigskip (i) The vacuum BD action in the JF is [1]%
\begin{equation}
S_{\text{JF}}=\frac{1}{16\pi }\int d^{4}x(-g)^{\frac{1}{2}}\left[ \varphi
\mathbf{R}+\varphi ^{-1}\omega g^{\mu \nu }\varphi _{,\mu }\varphi _{,\nu }%
\right] .
\end{equation}%
The field equations are%
\begin{eqnarray}
\square ^{2}\varphi &=& 0, \\
\mathbf{R}_{\mu\nu}-\frac{1}{2}g_{\mu\nu}\mathbf{R}&=&-\frac{\omega }{%
\varphi ^{2}}\left[ \varphi _{,\mu }\varphi _{,\nu }-\frac{1}{2}g_{\mu \nu
}\varphi _{,\sigma }\varphi ^{,\sigma }\right] \nonumber\\
&&-\frac{1}{\varphi }\left[
\varphi _{;\mu }\varphi _{;\nu }-g_{\mu \nu }\square ^{2}\varphi \right] ,
\end{eqnarray}
where $\square ^{2}\equiv (\varphi ^{;\rho })_{;\rho }$ and $\omega $ is a
dimensionless constant coupling parameter. The general solution, in
isotropic coordinates ($t,r,\theta ,\varphi $), is given by%
\begin{eqnarray}
d\tau ^{2}=g_{\mu \nu }dx^{\mu }dx^{\nu }&=&-e^{2\alpha (r)}dt^{2}+e^{2\beta
(r)}dr^{2} \nonumber\\
&&+e^{2\nu (r)}r^{2}(d\theta ^{2}+\sin ^{2}\theta d\psi ^{2}).
\end{eqnarray}%
Brans class I solution [2,3] correspond to the gauge $\beta -\nu =0$ and is
given by
\begin{eqnarray}
e^{\alpha (r)}&=&e^{\alpha _{0}}\left[ \frac{1-B/r}{1+B/r}\right] ^{\frac{1}{%
\lambda }}, \\
e^{\beta (r)}&=&e^{\beta _{0}}\left[ 1+B/r\right] ^{2}\left[ \frac{1-B/r}{1+B/r%
}\right] ^{\frac{\lambda -C-1}{\lambda }}, \\
\varphi (r)&=&\varphi _{0}\left[ \frac{1-B/r}{1+B/r}\right] ^{\frac{C}{\lambda
}}, \\
\lambda ^{2}&\equiv & (C+1)^{2}-C\left( 1-\frac{\omega C}{2}\right) >0,
\end{eqnarray}%

where $\alpha _{0}$, $\beta _{0}$, $B$, $C$, and $\varphi _{0}$ are
constants. The constants $\alpha _{0}$ and $\beta _{0}$ are determined by
asymptotic flatness condition as $\alpha _{0}=$ $\beta _{0}=0$.

(ii) Under the conformal transformation (often called Dicke transformations
[4])%
\begin{equation}
\widetilde{g}_{\mu \nu }=pg_{\mu \nu },\text{ \ \ }p=\frac{1}{16\pi }\varphi
,
\end{equation}%
and a redefinition of the BD scalar%
\begin{equation}
d\phi =\left( \frac{\omega +3/2}{\kappa }\right) ^{1/2}\frac{d\varphi }{%
\varphi },
\end{equation}%
the action (1) in the EF ($\widetilde{g}_{\mu \nu },\phi $) becomes%
\begin{equation}
S_{\text{EF}}=\int d^{4}x(-\widetilde{g})^{1/2}\left[ \widetilde{\mathbf{R}}%
+\kappa \widetilde{g}^{\mu \nu }\phi _{,\mu }\phi _{,\nu }\right]
\end{equation}%
leading to the field equations
\begin{eqnarray}
\widetilde{\mathbf{R}}_{\mu \nu } &=&-\kappa \phi _{,\mu }\phi _{,\nu } \\
\square ^{2}\phi  &=&0.
\end{eqnarray}%
The arbitrary constant $\kappa $ can have any sign $\pm 1$ but we chose $%
\kappa =+1$ in order to ensure that the stress of the source scalar field $%
\phi $ satisfies the energy conditions. The solutions of Eqs. (12) and (13)
can be obtained, using the transformations () and (). Redefining $B=\frac{m}{%
2}$, we then get what is called the Buchdahl solution [21,22]\footnote{%
It is also called Fisher solution [23] or Bergman-Leipnik solution [24]
caused by normal matter field but the solutions was expressed in a
non-transparent form. We are indebted to Prof. K.A. Bronnikov for pointing
it out. }%
\begin{eqnarray}
d\tau ^{2} &=&-\left( 1+\frac{m}{2r}\right) ^{-2\gamma }\left( 1-\frac{m}{2r}%
\right) ^{2\gamma }dt^{2}  \nonumber \\
&&+\left( 1+\frac{m}{2r}\right) ^{2(1+\gamma )}\left( 1-\frac{m}{2r}\right)
^{2(1-\gamma )}\times   \nonumber \\
&&\left[ dr^{2}+r^{2}d\Omega ^{2}\right] , \\
\phi (r)&=&\sqrt{\left( \omega +\frac{3}{2}\right) \left( \frac{C^{2}}{\lambda
^{2}}\right) }\ln {\left[ \frac{1-\frac{m}{2r}}{1+\frac{m}{2r}}\right] }. \\
\gamma &=&\frac{1}{\lambda }\left( 1+\frac{C}{2}\right) .
\end{eqnarray}%
With this $\gamma $, and using (8), the scalar field can be re-expressed as%
\begin{equation}
\phi (r)=\sqrt{2(1-\gamma ^{2})}\ln {\left[ \frac{1-\frac{m}{2r}}{1+\frac{m}{%
2r}}\right] .}
\end{equation}%
The solution (14,17) is now completely characterized by a single free
parameter $\gamma $, a constant of integration of GR equations (12,13) that
no longer has anything to do with the Machian parameter $\omega $.

However, in the limit $\omega \rightarrow \infty $, one should recover GR
from Brans-Dicke theory\footnote{%
The limit $\omega \rightarrow \infty $ is not as simple as it looks. The
passage to GR in that limit is said to be anomalous and is still being
discussed (see e.g., [25-29]), but we disregard these issues here. Most
recently, Faraoni and C\^{o}t\'{e} [29] have shown that the symmetry is
preserved even in the limit $\omega \rightarrow \infty $.}. But the Buchdahl
solution self-consistently satisfies the EF equations (12,13), where the
metric (14) and the scalar field $\phi $ in Eq.(17) have no $\omega $
dependence and the limit $\omega \rightarrow \infty $ has no meaning.
However, the intermediate expression (15) for $\phi $ does contain $\omega $
and in that case, one might hope to talk of the limit $\omega \rightarrow
\infty $. For this limit in EF to be meaningful, the WFL observables from
both the metrics (4) and (14) should exactly coincide in this limit. To show
that this is indeed the case, one then has to use the previous incarnation
of $\gamma $ connecting it to $\omega $.

Using the weak field expression for
\begin{equation}
C=-\frac{1}{2+\omega }
\end{equation}%
in $\lambda $ in (8) and $\gamma $ in (16), it follows that
\begin{equation}
\gamma =\lambda =\sqrt{\frac{3+2\omega }{4+2\omega }},
\end{equation}%
which shows $\gamma \rightarrow 1$ as $\omega \rightarrow \infty $. The
expansion of JF metric component.(5) yields%
\begin{equation}
e^{2\alpha (r)}\simeq 1-\frac{4B}{\lambda r}\simeq 1-\frac{2M}{r}
\end{equation}%
from which one obtains the Schwarzschild mass $M_{\text{Sch}}^{\text{JF}}=%
\frac{2B}{\lambda }$, and similarly from the EF metric component in (14), $%
M_{\text{Sch}}^{\text{EF}}=m\gamma $. The ADM mass for JF metric (4) is $M_{%
\text{ADM}}^{\text{JF}}=\frac{2B}{\lambda }(C+1)=M_{\text{Sch}}^{\text{JF}%
}(C+1)$. Similarly, for EF metric (14), the mass is $M_{\text{ADM}}^{\text{EF%
}}=m\gamma =M_{\text{Sch}}^{\text{EF}}$ and we already see that the
gravitating masses are different in JF and EF. Thus the frames are
qualitatively distinguishable already showing the inequivalence of the JF
and EF though all the masses coincide at $\omega \rightarrow \infty $.
However, we want to know how the experimental observables quantitatively
differ between them for \textit{finite} $\omega $, for which we turn to WFL
observables.

\section{WFL observables}
Gao et al [19] have extended Keeton-Petters method [30] up to 4th order,
which essentially are the PPN coefficients of the light deflection angle
(for some useful recent applications of the method, see [31,32]). For the
vacuum BD\ solution in the JF, the coefficients are (we quote Gao's
coefficients up to third order without any loss of rigor for our argument):

\begin{eqnarray}
A_{1}^{\text{JF}} &=&\frac{6+4\omega }{2+\omega }, \\
A_{2}^{\text{JF}} &=&\frac{\pi (30\omega ^{2}+89\omega +66)}{8(2+\omega )^{2}%
}, \\
A_{3}^{\text{JF}} &=&\frac{2(3+2\omega )^{2}\pi (30\omega ^{2}+89\omega +66)%
}{8(2+\omega )^{2}}.
\end{eqnarray}%
For the Buchdahl solution of GR, the PPN coefficients are expressed in terms
of $\gamma $. When expressed in terms of $\omega $ up to third order, they
yield
\begin{eqnarray}
A_{1}^{\text{EF}} &=&4\;, \\
A_{2}^{\text{EF}} &=&\pi \left( 4-\frac{1}{4\gamma ^{2}}\right) =\frac{\pi
\left( 22+15\omega \right) }{6+4\omega }\;, \\
A_{3}^{\text{EF}} &=&16\left( 3-\frac{1}{3\gamma ^{2}}\right) =\frac{%
368+256\omega }{9+6\omega }\;.
\end{eqnarray}%
They are evidently \textit{not} the same expressions derived by of Gao et al
[19] directly from the original vacuum Brans-Dicke theory in the JF.
However, as $\omega \rightarrow \infty $, we retrieve the vacuum GR values
$$A_{1}=4,\quad A_{2}=\frac{15\pi }{4},\quad A_{3}=\frac{128}{3},$$
and similarly following the method of Gao et al [19], we have verified in
the EF that $A_{4}=\frac{3465\pi }{64}$ in that limit. These are exactly the
same values obtained also from Eqs.() in the same $\omega \rightarrow \infty
$ limit. The arguments and calculations above illuminate the fundamental
difference between the lensing observables of Brans-Dicke theory and
Einstein's theory \textit{except} in the limit $\omega \rightarrow \infty $.
\

Since, after all, $\omega $ is not infinite but very large, it would be
curious to see how much accuracy in measurement would be required to
distinguish the WFL observables between JF and EF. A value of $\omega $
follows from the Cassini spacecraft experiment by Bertotti et al [20], which
we shal use. They measured the round trip time delay $\Delta t$ of light
between the ground antenna and the spacecraft, at distances $r_{1}$ and $%
r_{2}$ respectively from the Sun with mass $M_{\odot }$ using the model
independent Robertson Coefficient $\gamma ^{\text{RC}}$ of the centrally
symmetric gravity field [33] [Note: $\gamma ^{\text{RC}}$ has nothing to do
with the EF $\gamma $ in Eq.(16)]: \
\begin{equation}
\Delta t=2(1+\gamma ^{\text{RC}})\frac{GM_{\odot }}{c^{3}}\ln \left( \frac{%
4r_{1}r_{2}}{b^{2}}\right) ,
\end{equation}%
where $\left( r_{1},r_{2}\right) <<b$, the impact parameter. The value of $%
\gamma ^{\text{RC}}$ is $0$ in Newtonian gravity and $1$ in GR. The result
of the Cassini spacecraft experiment is $\gamma ^{\text{RC}}=1+\left( 2.1\pm
2.3\right) \times 10^{-5}$ [20]. In the BD theory, one has $\gamma ^{\text{RC%
}}=\frac{1+\omega }{2+\omega }$, which leads to a value $\left\vert \omega
\right\vert \sim 50,000.$

We consider the WFL observables to be the image position $\theta $ (scaled
by the Einstein angle $\mathfrak{v}_{E}$) and total magnification $\mu _{%
\text{tot}}$. Consider the expansion
\begin{equation}
\theta =\theta _{0}+\theta _{1}\varepsilon +\theta _{2}\varepsilon ^{2}+...\;,
\end{equation}%
where, for an arbitrary gravitating mass $M$, and the small expansion
parameter $\varepsilon $,
\begin{eqnarray}
\varepsilon &=&\frac{\theta _{\bullet }}{\mathfrak{v}_{E}}\;,\theta _{\bullet}=\tan ^{-1}\left( \frac{m_{\bullet }}{d_{OL}}\right) \;, \nonumber\\
m_{\bullet }&=&\frac{GM}{c^{2}},\;\mathfrak{v}_{E}=\left( \frac{4Md_{LS}}{d_{OL}d_{OS}}\right)^{1/2}\;,
\end{eqnarray}%
where $d_{OL}$ is the observer-lens distance, $d_{OS}$ is the
observer-source distance and $d_{LS}$ is the lens-source distance. We shall
consider the known black hole SgrA* as the lens and for its mass, the
correction $\theta _{1}\varepsilon $ is too small compared to the zeroth
order term $\theta _{0}$ in both frames as we will soon see. In the JF, the
coefficients are [19] (considering only the positive parity image):%
\begin{eqnarray}
\theta _{0}^{\text{JF}} &=&\frac{1}{2}\left[ \beta +\left( \sqrt{\beta^{2}+4\left( \frac{3+2\omega }{4+2\omega }\right) }\right) \right] , \\
\theta _{1}^{\text{JF}} &=&\frac{\pi (30\omega ^{2}+89\omega +66)}{16(\omega+2)[2\theta _{0}^{2}(2+\omega )+3+2\omega ]},
\end{eqnarray}%
Similarly, in the EF, we find that

\begin{eqnarray}
\theta _{0}^{\text{EF}} &=&\frac{1}{2}\left[ \beta \pm \left( \sqrt{\beta^{2}+A_{1}^{\text{EF}}}\right) \right] \nonumber \\
&=&\frac{1}{2}\left[ \beta +\left( \sqrt{\beta ^{2}+4}\right) \right] , \\
\theta _{1}^{\text{EF}} &=&\frac{A_{2}^{\text{EF}}}{A_{1}^{\text{EF}}+4\left( \theta _{0}^{\text{EF}}\right) ^{2}} \nonumber \\
&=&\frac{\pi \left( 22+15\omega \right) }{\left( 6+4\omega \right) \left[ 4+\left( \beta +\sqrt{\beta ^{2}+4}\right) ^{2}\right] },
\end{eqnarray}%
Technically, it is difficult to measure individual magnifications of images,
so we consider the total magnification, which are [19]

\begin{eqnarray}
\mu _{0,\text{tot}}^{\text{JF}} &=&\frac{A_{1}^{\text{JF}}+2\beta ^{2}}{%
2\beta \sqrt{A_{1}^{\text{JF}}+\beta ^{2}}}=\frac{\beta ^{2}(\omega
+2)+3+2\omega }{\beta (\omega +2)n}, \\
\mu _{1,\text{tot}}^{\text{JF}} &=&-\frac{A_{2}^{\text{JF}}}{2\left( A_{1}^{%
\text{JF}}+\beta ^{2}\right) ^{3/2}} \nonumber \\
&=&-\frac{\pi (3+2\omega )(15\omega +22)n}{%
16[\beta ^{2}(2+\omega )+4\omega +6]^{2}}, \\
n &=&\sqrt{\beta ^{2}-\frac{2}{2+\omega }+4}.
\end{eqnarray}%
Similarly, in the EF,

\begin{eqnarray}
\mu _{0,\text{tot}}^{\text{EF}} &=&\frac{A_{1}^{\text{EF}}+2\beta ^{2}}{%
2\beta \sqrt{A_{1}^{\text{EF}}+\beta ^{2}}}=\frac{2+\beta ^{2}}{\beta \sqrt{%
4+\beta ^{2}}} \\
\mu _{1,\text{tot}}^{\text{EF}} &=&-\frac{A_{2}^{\text{EF}}}{2\left( A_{1}^{%
\text{EF}}+\beta ^{2}\right) ^{3/2}}=-\frac{\pi (22+15\omega )}{2(6+4\omega
)\left( 4+\beta ^{2}\right) ^{3/2}}
\end{eqnarray}%
where we have used $A_{1}^{\text{EF}}=4$.

\section{Accuracies needed}
To get a flavor of the accuracy needed, we take, for illustration, $\beta
=0.1$ and following the relevant values for the lens SgrA* [30] with $%
d_{LS}=10$ parsec, we take the small expansion parameter to be $\varepsilon
=1.3\times 10^{-4}$, the Einstein angle $\mathfrak{v}_{E}=0.068$ arcsec, and
from the Cassini spacecraft experiment, $\omega =50000$ [20]. Restoring the
dimensions by multiplying both sides of the series by $\mathfrak{v}_{E}$, we
get
\begin{eqnarray}
\theta _{0}^{\text{JF}}\mathfrak{v}_{E} &=&71484.94\text{ }\mu \text{arcsec}\;,\nonumber \\
\theta _{1}^{\text{JF}}\varepsilon \mathfrak{v}_{E}&=&12.3678\text{ }%
\mu \text{arcsec} \\
\Rightarrow \theta _{\text{tot}}^{\text{JF}}&=&\theta _{0}^{\text{JF}}%
\mathfrak{v}_{E}+\theta _{1}^{\text{JF}}\varepsilon \mathfrak{v}%
_{E}=71497.314\text{ }\mu \text{arcsec} \\
\theta _{0}^{\text{EF}}\mathfrak{v}_{E} &=&71484.6\text{ }\mu \text{arcsec}\;, \nonumber \\
\theta _{1}^{\text{EF}}\varepsilon \mathfrak{v}_{E}&=&12.3678\text{ }%
\mu \text{arcsec} \\
\Rightarrow \theta _{\text{tot}}^{\text{EF}}&=&\theta _{0}^{\text{EF}}%
\mathfrak{v}_{E}+\text{ }\theta _{1}^{\text{EF}}\varepsilon \mathfrak{v}%
_{E}=71496.975\text{ }\mu \text{arcsec}
\end{eqnarray}%
The qualitative conclusion is that the image angular positions are closer to
the optical axis $OL$ \ in the EF than in the JF. Evidently, the first order
corrections are too small in both frames compared to the zeroth order, $%
\theta _{1}^{\text{JF,EF}}\varepsilon \mathfrak{v}_{E}<<\theta _{0}^{\text{%
JF,EF}}\mathfrak{v}_{E}$ and higher order terms would be even smaller, hence
ignored. Looking at the total angular position, $\theta _{\text{tot}}^{\text{%
JF}}$ and $\theta _{\text{tot}}^{\text{JF}}$, it is clear that future
astrometric missions would need to attain an angular measurement accuracy
better that $0.3$ $\mu \text{arcsec}$ to observationally distinguish between
the two frames. This is certainly a challenging task but not impossible in
the future. Likewise
\begin{eqnarray}
\mu _{0,\text{tot}}^{\text{JF}} &=&10.0374,\text{ }\mu _{1,\text{tot}}^{%
\text{JF}}=-0.733554 \\
\mu _{\text{tot}}^{\text{JF}} &=&9.30386 \\
\mu _{0,\text{tot}}^{\text{EF}} &=&10.0375,\text{ }\mu _{1,\text{tot}}^{%
\text{EF}}=-0.733558, \\
\mu _{\text{tot}}^{\text{EF}} &=&9.30390, \\
&\Rightarrow &\mu _{\text{tot}}^{\text{JF}}/\mu _{\text{tot}}^{\text{EF}%
}=1-5\times 10^{-6}.
\end{eqnarray}%
This shows that, qualitatively, the image magnification is more in the EF
than in the JF. In practice, one measures the\textit{\ magnification factor }%
[34], which seems a bit easier to observe, to which we devote the next
section.

\section{Magnification factor}
\bigskip We start with the leading order angular positions $\theta _{\pm }$
of the images dropping the subscript $0$:
\begin{equation}
\theta _{\pm }=\frac{1}{2}\left[ \beta \pm \left( \sqrt{\beta ^{2}+A_{1}}%
\right) \right] ,
\end{equation}%
where $A_{1}$ could be giv$A_{1}^{\text{JF}}$ or $A_{1}^{\text{EF}}=4$ and
the angles for both parity images $\theta _{\pm }$ and $\beta $ are as
before scaled by the Einstein angle $\mathfrak{v}_{E}$. The polar width $%
\Delta \theta _{\pm }$ of the images change and its magnitude can be
obtained by differentiating Eq.(63):
\begin{equation}
\Delta \theta _{\pm }=\left( \frac{1}{2}\right) \left[ 1\pm \frac{\beta }{%
\sqrt{\beta ^{2}+A_{1}}}\right] \Delta \beta .
\end{equation}%
Since the angular width of the source $\Delta \beta \neq 0$, this result
implies a distorted and elongated shape of the images that have been
confirmed by observations.

The ratio of the brighness of the individual images $\mu _{\pm }$ to the
unlensed brightness $\mu _{\ast }$ at the angular positions $\theta _{\pm }$
is given by the individual magnifications

\begin{eqnarray}
\frac{\mu _{\pm }}{\mu _{\ast }} &=&\left\vert \left( \frac{\theta _{\pm }}{%
\beta }\right) \left( \frac{d\theta _{\pm }}{d\beta }\right) \right\vert
\nonumber \\
&=&\frac{1}{4}\left[ \frac{\beta }{\sqrt{\beta ^{2}+A_{1}}}+\frac{\sqrt{%
\beta ^{2}+A_{1}}}{\beta }\pm 2\right] .
\end{eqnarray}%
We can draw a very interesting conclusion from here: Since $x+1/x\geq 2$, we
can conclude that, for $\beta <0$, the brightness $\left\vert \mu
_{-}\right\vert >\mu _{+}$ showing that the image $\theta _{-}$ is brighter
than the image $\theta _{+}$. However, the magnitudes of individual image
brightness $\mu _{\pm }$ for any given source do not differ very greatly
thus making the individual measurements difficult. In this case, another
very useful quantity is the total magnification over the background $\mu
_{\ast }$, called the \textit{magnification factor} $\mathcal{F}$ defined by
(see for details, Hartle [34])
\begin{eqnarray}
\mathcal{F}^{\text{JF}} &=&\left. \frac{\mu _{\text{tot}}}{\mu _{\ast }}%
\right\vert _{\text{JF}}=\frac{\mu _{+}-\mu _{-}}{\mu _{\ast }}=\frac{1}{2}%
\left[ \frac{\beta }{\sqrt{\beta ^{2}+A_{1}}}+\frac{\sqrt{\beta ^{2}+A_{1}}}{%
\beta }\right]   \nonumber \\
&=&\frac{1}{2}\left[ \frac{\beta }{\sqrt{\beta ^{2}+4\left( \frac{3+2\omega
}{4+2\omega }\right) }}+\frac{\sqrt{\beta ^{2}+4\left( \frac{3+2\omega }{%
4+2\omega }\right) }}{\beta }\right] ,
\end{eqnarray}%
the ratio being always greater than unity. For GR, $A_{1}=4$ and so
\begin{equation}
\mathcal{F}^{\text{EF}}=\left. \frac{\mu _{\text{tot}}}{\mu _{\ast }}%
\right\vert _{\text{EF}}=\frac{\mu _{+}-\mu _{-}}{\mu _{\ast }}=\frac{1}{2}%
\left[ \frac{\beta }{\sqrt{\beta ^{2}+4}}+\frac{\sqrt{\beta ^{2}+4}}{\beta }%
\right] .
\end{equation}

The best result is obtained when $\beta $ is small, that is, when the source
is very close to the optical axis $OL$, the total magnification factor could
be quite large that should be observable. Accordingly, we take $\beta =0.1$
and translate the limit on $\omega =50000$ from the Cassini spececraft
experiment [20] into image magnification factor. Then we find that%
\begin{equation}
\frac{\mathcal{F}^{\text{JF}}}{\mathcal{F}^{\text{EF}}}=1-4.9\times 10^{-6},
\end{equation}%
which is expectedly almost the same as in Eq.(47). Recent measurements of
gravitationally lensed quasars provide total magnifications with too wide
error margins and the factor ratio is impossible to measure [35].
Measurements of image positions could probably be easier provided an
accurary at the level of fractional $\mu $arcsec is attained, challenging
but not impossible.

\section{Conclusions}
The answer to the question posed in the title is yes. We considered the
vacuum Brans class I solution (4-8) and its conformal transform to the
Buchdahl solution (14,17), both solutions being self consistent and physical
as defined in the introduction. It is evident that these solutions together
with their parent theories, vacuum BD and GR respectively, can very well
exist independently of knowing each other. In this case, the constants ($%
C,\lambda ,\omega )$ from the solutions (4-8) and $\gamma $ from (14,17) are
unconnected since $\omega $ has no meaning in the EF. But if one believes
that the matter scalar field $\phi $ in the EF\ still has a \textit{memory}
of its previous incarnation of $0-$component gravity in the JF, then the
connection Eq.(16) has a meaning and using it, one can easily retrieve the
JF metric from the EF metric and vice versa. However, nobody measures the
metric but only the experimental observables, which we have chosen to be the
WFL observables.

In general, reformulation of any scalar-tensor theory in a new conformal
frame leads to anomalous coupling to matter making the reformulation a
physically inequivalent theory and it is believed that experiments involving
motion of massive timelike particles would reveal this difference [10]. We
showed here that quantitatively different results appear for \textit{finite}
$\omega $ even when the anomalous coupling is absent.

Recent derivations by Gao et al [19] of \ the WFL obervables in the JF for $%
C(\omega )=-\frac{1}{2+\omega }$ have given us an opportunity to exactly
compare their results with those obtained in the EF. Using the expression
for $\gamma (\omega )$ of Eq.(19) in the EF observables, we argue that one%
\textit{\ cannot} retrieve the values of WFL observables in the JF. As
illustration, we chose the black hole SgrA* as the lens and used the value $%
\omega =50000$ obtained in the Cassini spacecraft experiment [20] to show
that angular measurement accuracy better that $0.3$ $\mu \text{arcsec}$ is
required to observationally distinguish between the two frames. In the same
way, the image magnification would differ in the two frames by a factor $%
\left( 1-4.9\times 10^{-6}\right) $. The accuracy needed in both cases is
challenging but may not be impossible to attain in the future.

\section*{Aknowledgements}

The reported study was funded by RFBR according to the research project No. 18-32-00377.

\end{document}